# Electron-beam assisted selective growth of graphenic carbon thin films on $SiO_2$/Si and quartz substrates

M.A. Knyazev, D.M. Sedlovets[*], O.V. Trofimov, A.N. Redkin

*Institute of Microelectronics Technology and High-purity Materials, Russian Academy of Science, Chernogolovka, Moscow District, 6 Academician Ossipyan str., 142432 Russia.*

**Abstract:**

The first selective growth of graphenic carbon thin films on silicon dioxide is reported. A preliminary e-beam exposure of the substrate is found to strongly affect the process of such films growth. The emphasis is placed on the influence of substrate exposure on the rate of carbon deposition. The explanation of this effect is proposed. The data of electrical and optical measurements and the results of atomic force and scanning electron microscopy and Raman spectroscopy studies are reported. The results suggest that the selective growth of graphenic carbon thin films on an irradiated $SiO_2$/Si substrate is a promising approach to producing a microstructure at the pre-synthesis step without full lithography process.

## 1. Introduction

After the discovery of graphene, the microelectronics technology faced a difficult challenge to develop the efficient deposition methods for one- and few-layer films. Of particular interest are the direct (transfer-free) synthetic techniques, because the transfer process from metal surfaces is cumbersome, resource intensive and is likely to cause the formation of wrinkles and defects.

[*] Corresponding author: tel.: +7(49652)441-90; e-mail: sedlovets@iptm.ru

The first work in this direction was focused on the possibility of direct deposition of a thin film composed of graphene grains over magnesium oxide surface through the pyrolysis of unsaturated hydrocarbons [1]. It was followed by a number of works devoted to the non-catalytic growth of such films by carbon sputtering on monocrystalline Si [2] and mica [3]. Direct synthesis of graphenic carbon thin films on various surfaces (glass, mica, quartz, sapphire, Si, SiC, $SiO_2$) was carried out by low-temperature ($550^{o}C$) methane pyrolysis using the plasma enhanced chemical vapor deposition (CVD) [4]. In work [5] low-temperature decomposition of methane was performed in the electron cyclotron resonance plasma. The classic method of graphene CVD-growth, high-temperature ($1000^{o}C$ and above) pyrolysis of a methane-hydrogen mixture, is also used for synthesis on non-metallic surfaces, only with a longer deposition time - up to 3 hours [6]. The authors of [7] showed that polycrystalline few-layer graphene can be directly obtained on the $SiO_2$ surface by this technique if synthesized samples were pre-calcined in flowing air at 800°C. Earlier, we also reported the method of obtaining graphenic carbon thin films by pyrolysis of ethanol vapor [8]. This is a simple, low-cost and scalable approach, which excludes the use of flammable gases. At high temperatures (900-1000$^{o}$C), the pyrolysis of ethanol allows obtaining carbon films not only on a catalytic metal surface, but also on dielectric substrates, such as quartz and $SiO_2$/Si. The obtained material is a transparent conductive continuous film composed of randomly oriented, interconnected graphene grains with an average crystallite size about 20 nm. In scientific literature such films are often named

«nanographene» [1,4] or «graphene-like» [5]. We use the term "graphenic" [9], i.e. $sp^2$ bonded carbon, for the films consisted of graphene flakes.

In spite of a large number of works on transfer-free carbon deposition, they primarily deal with non-selective deposition studies. For microstructuring of synthesized films usually requires the lithography process, which can however destroy the film texture [10], especially because it has a low structural perfection. Therefore, direct producing of a microstructure at the pre-synthesis step is of great value for technological applications. In this paper we present, for the first time, an unprecedented example of selective growth of graphenic carbon thin films on $SiO_2$/Si, the basic microelectronics substrate, which allowed producing a microstructure at the pre-synthesis step.

The carbon deposition mechanism is discussed below in relation to the role of the surface substrate condition on the nanocrystalline carbon films growth.

The ordering of carbon atoms in a hexagonal lattice is a complex physicochemical phenomenon and its detailed mechanism is not yet understood. The authors of [11] proposed two mechanisms of process. The first is a self-assembly of graphene flakes resulting from the pyrolysis of $CH_4$, viz. graphene nuclei formed in the gas phase lay over and coalesce on the thermodynamically stable substrate. The other mechanism of the CVD of graphene-based material on $SiO_2$ might be of catalytic nature, i.e. the formation and decomposition of surface carbide intermediates. On the other hand, it was found in [5] that the process is non-catalytic, since it is known that the catalytic dissociation of hydrocarbons on the $SiO_2$ surface is accompanied by the formation of the intermediate silicon carbide, which in this case does not occur. Presumably, another growth mechanism can operate here, namely, adsorption of hydrocarbon radicals on the surface, followed by graphitization. As shown in [12], direct graphene growth on sapphire substrates does not undergo lateral growth on the substrates. It seems that the graphene grains just fall onto the substrates, and the limited lateral growth procedure may only take place when the flakes overlap.

It is appears that in all the cases the state of the surface is one of the factors influencing the process of synthesis. Therefore, the substrate modification may be one of the ways to control the graphene-based materials deposition. In the present work we describe the influence of electron beam (e-beam) irradiation of the $SiO_2$-layer on the process of graphenic carbon thin films obtaining.

## 2. Material and methods

### 2.1. A marker structure fabricating

$SiO_2$/Si (300 nm of thermally grown $SiO_2$ on a Si wafer) and quartz plates were used as a substrate for our samples. First a marker structure was designed and fabricated to determine e-beam exposed areas. The e-beam lithography (EBL) and lift-off process were used for the structures fabricating. A laboratory lithograph was used for the EBL.

It is based on a scanning electron microscope (SEM) ZEISS EVO-50 and the hardware-software system Nanomaker. The exposure data was prepared in the Nanomaker software part. After e-beam exposure and electron resist development a 80 nm Ni-film was deposited on the samples using thermal sputtering. This technological step was made by vacuum evaporation. The next step was a lift-off process and the obtained marker structure is imaged in Figure 1. The marker structure had ten 150x150 μm squares which are separated by Ni stripes.

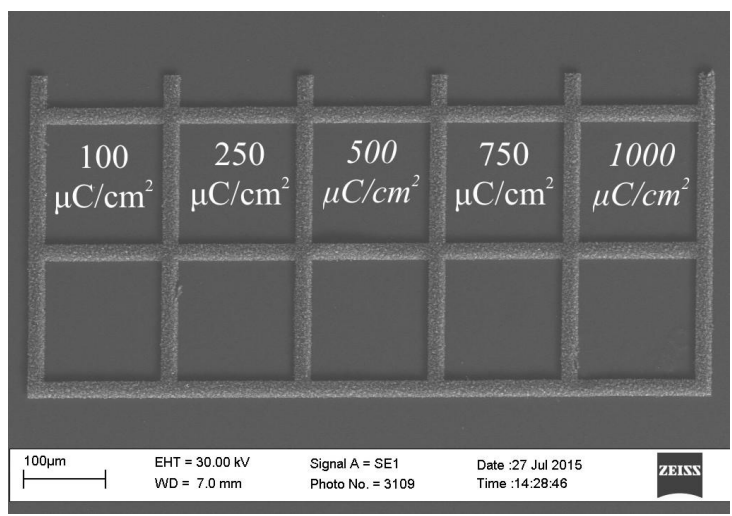

Figure 1. SEM image of the nickel marker structure.

**2.2. E-beam irradiating of silicon dioxide**

A laboratory lithograph was used for irradiating of silicon dioxide. The squares were exposed with the energy of the irradiating electrons equal to 5keV. The exposure doses in five upper areas were varied from 100 to 1000 μC/cm$^2$. Different test structures (grids and circles) in five lower squares were exposed with dose equal to 1000 μC/cm$^2$.

**2.2. Growth of graphenic carbon thin films**

Carbon films were grown in a flow-type quartz reactor incorporated in a CVD apparatus, which included a tube furnace, a temperature control system, a reagent (ethanol and water) feed system and a vacuum pump. The synthesis was carried out in an inert gas (argon) flow at a reduced pressure (~10$^3$ Pa). The feedstock flow rate was controlled by a peristaltic pump and was varied from 2 to 4 ml/h. The water-alcohol mixture (96% ethanol and distilled water) was fed directly into the evaporation zone to allow the consistency between the composition of the feed solution and the ratio of the components of the gas phase. After the deposition was completed, the reagent feeding was stopped and the reactor was cooled down to room temperature in an inert gas flow.

The synthesis temperature was 950$^{o}$C. The reaction times and the water content in the initial mix varied from 1 to 30 min and from 0 to 50%, respectively.

**2.3. Materials characterization**

SEM images were acquired with a SEM ZEISS Evo 50 system.

The thickness of the carbon films was estimated from the surface height difference at the film/substrate interface using an atomic force microscope (AFM) (home-made IMT RAS).

Raman spectra were taken with a Bruker Senterra micro-Raman system equipped with a 532 nm laser in the 300-3700 cm$^{-1}$ range under 10 cm$^{-1}$ resolution.

Optical spectra of the films were recorded on a Bruker Vertex 70V spectrometer combined with a HYPERION 2000 IR-microscope in the 500-600 nm range.

The resistance was measured directly between the two contacts taken from the top of the selectively grown films. For this purpose, two nickel strips were deposited by vacuum evaporation.

Electrical and optical measurements were performed on films deposited on quartz substrate. Other studies were carried out on films deposited on $SiO_2$/Si substrate.

**3. Results and discussion**

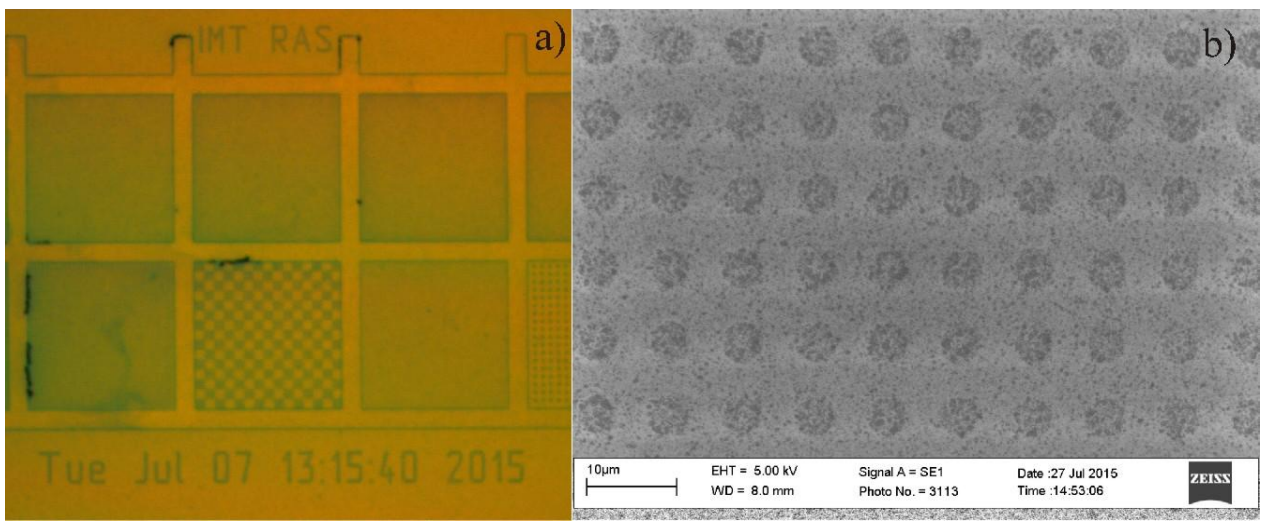

Figure 2. Optical (a) and SEM (b) images of the graphenic carbon thin film on the SiO$_2$/Si pre-exposed substrate. Synthesis temperature was 950$^o$C, deposition time was 30 min.

The graphene-based film was grown on the SiO$_2$/Si pre-exposed substrate through the pyrolysis of ethanol vapor at 950$^o$C for 30 min. The Figure 2a shows an optical image of the obtained sample. The darker areas, namely the lettering «IMT RAS», the date and the grids inside the squares are exposed by an electron beam before synthesis. Noteworthy that under such synthesis conditions the film deposits everywhere on the SiO$_2$/Si, but Figure 2a shows that the difference exists in color between the un- and exposed sectors. As follows from Figure 2b, graphene flakes (black dots) cover all the substrate, wherein their highest density is observed in the exposed circles. The AFM studies revealed that the carbon films on the irradiated areas are thicker. This observation was supported by the data of optical measurements for the films obtained on quartz substrate.

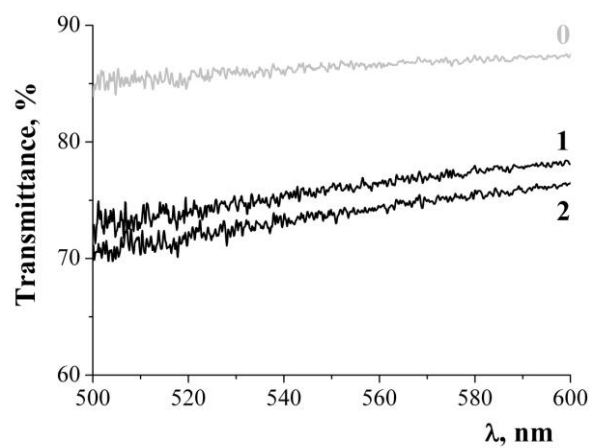

Figure 3. Optical spectra of the sample prepared at 950°C on quartz. Deposition time was 20 min. Exposure doses: 0 – unexposed. 1- 100 μC/cm$^2$; 2 - 1000 μC/cm$^2$.

Optical transmittance of the films obtained on exposed and unexposed sectors of quartz substrate is shown in Figure 3. It is seen that in the 500-600 nm range the films on irradiated region (spectra 1,2) become «darker» then on the remaining surface (spectra 0). This means that film thickness increases. Thereby, the rates of graphenic carbon thin films deposition on the e-beam-exposed sites and on the unexposed areas are different.
So, it is possible to assume that synthesis conditions can be chosen such that the film be deposited only on the exposed sites while on the remaining areas the film would have no time to form. There propose several different ways to do this: decreasing the synthesis temperature, lessening the deposition time and addition of water to the reagent. The last variant is supported by the fact that earlier we reported that the addition of water led to a decrease in the film thickness with a simultaneous slight improvement in the quality of the resulting carbon films [13]. Decreasing of the synthesis temperature is not advisable because the obtained films have worsened properties. Changes only in synthesis time lead to that a film is deposited either everywhere or nowhere. However, the addition of water to the initial mix together with simultaneous reduction of time gave to a successful result.

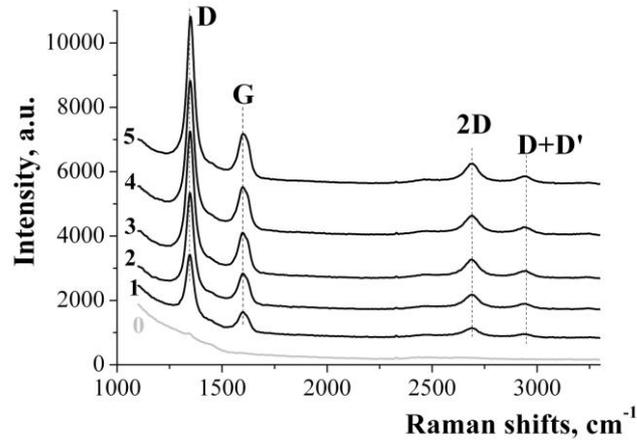

Figure 4. Raman spectra of the sample prepared at 950°C on $SiO_2$/Si. Deposition time was 10 min, an initial mix contained 50% $H_2O$. Exposure doses: 0 – unexposed. 1- 100 µC/cm$^2$; 2 – 250 µC/cm$^2$; 3 – 500 µC/cm$^2$; 4 – 750 µC/cm$^2$; 5 – 1000 µC/cm$^2$.

The conditions for the selective growth were found. Figure 4 shows the Raman spectra for the sample prepared by the pyrolysis of ethanol-water mixture (50% $H_2O$) at 950°C for 10 min. The spectrum 0 was received from unexposed region and the spectra 1-5 were received from areas with different doses of preliminary e-beam exposure. It is seen that the carbon film is present only on the exposed sites (spectra 1-5), while other areas remain clean (spectrum 0). The D band is observed at ~1350 cm$^{-1}$, suggesting the presence of disordered structural defects (e.g., amorphous carbon or edges that break the symmetry and selection rule). The G band at ~1600 cm$^{-1}$ arises from the vibration of sp$^2$ bonded carbon atoms. The 2D band at ~2690 cm$^{-1}$, originates from the second order double resonant Raman scattering [14]. The D+D' (~2950 cm$^{-1}$) is the combination of phonons with different momenta, a defect requires for its activation [15]. Note that Raman spectra of the samples obtained on a non-metal surface have some specific features: an intense D-, a broaden G-, and a weak and broad 2D-peak, as well D+D'-peak appears [1,4,5,13].

The sample was also investigated by AFM and an average difference in the surface heights at the film/substrate interface equal to ~4 nm was observed (Figure 5). Some particles detected on the surface (see fig. 5), are most likely due to the co-deposition of nanographite during growth as described in [11].

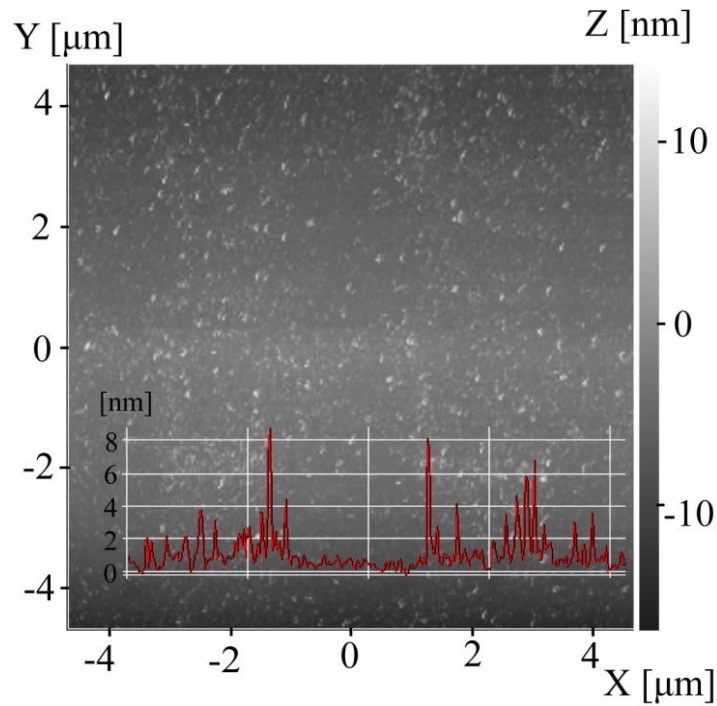

Figure 5. AFM image of the selectively grown graphenic carbon thin film. Insert on figure shows the surface heights difference at the film/substrate interface.

For electrical measurements a graphenic carbon stripe between the two Ni electrodes was grown. A length of the stripe was 1 mm and a width – 0,3 mm. Further, current-voltage (I-U) characteristic was measured (Figure 6). Linear (ohmic) behavior was observed. The sheet resistance obtained in the measurements was ~10 kΩ/square. So, the selectively grown film is conductive and, accordingly, continuous. The present result confirms that the proposed method of graphenic films obtaining is a promising approach to producing of microstructures at preliminary production step, without post-synthesis lithography.

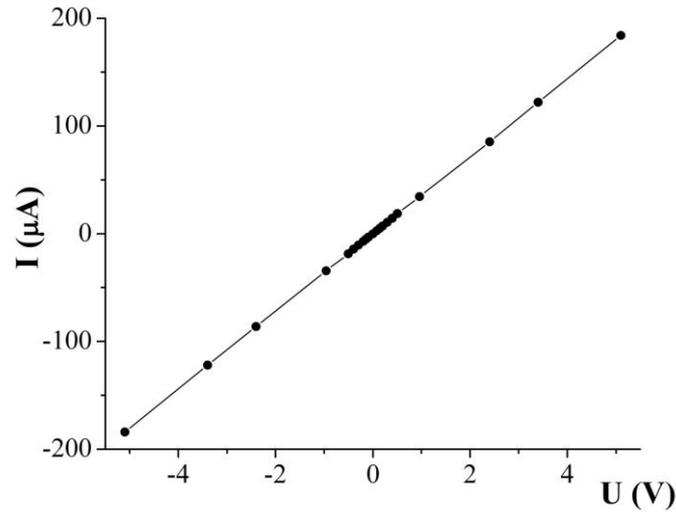

Figure 6. I-U curve of the graphenic carbon stripe (1 mm x 0,3 mm) which was selectively grown between the two Ni electrodes on quartz.

A few probable mechanisms of electrons exposure influence on the material growth can be proposed. Firstly, it is known that in SEM carbon contamination occurs on the surface during electron exposure [16]. Therefore contaminants can act as growth nuclei during synthesis. An other mechanism consists in possible substrate surface modification by e-beam irradiation to give rise to intermediate carbides, nanoclusters or defects [17]. At lastly, e-beam exposure can cause charging of dielectric films and part of the charge remain in the film after exposure [18-20] and can also affect the deposition process.

Special measures were taken to exclude the possibility of surface contamination by carbon particles resulting from e-beam exposure which can act as growth nuclei. For which purpose the following experiment was carried out. Before exposure a thin layer (20 nm) of aluminum was deposited on a $SiO_2$ substrate, the sample was subjected to e-beam irradiation and then the metal was removed by liquid etching. The above procedure protected the surface of silicon dioxide from possible formation of carbon nucleation centers. In spite of these measures, difference in the film thickness on exposed and unexposed areas did occur. The obtained samples were calcined in air at 700$^o$C for 30 minutes in order to burn off all the carbon. Then the synthesis was carried out again. And even then, the influence of the preliminary substrate e-beam exposure on the graphenic carbon deposition rate was apparent.

Further, the formation of intermediate carbides can be directly excluded because even at extremely short synthesis time (5 minutes or less) the Raman spectra exhibited no signals from SiC. Nanocluster or defect formation on the substrate surface is also unlikely, because the energy which the electrons exert the $SiO_2$-layer is too small to "shift" atoms. Moreover, AFM investigations revealed no change in the topography (Figure 7a) of the $SiO_2$ surface after e-beam irradiation. There was no detectable relief of the surface which could act as nucleation centers.

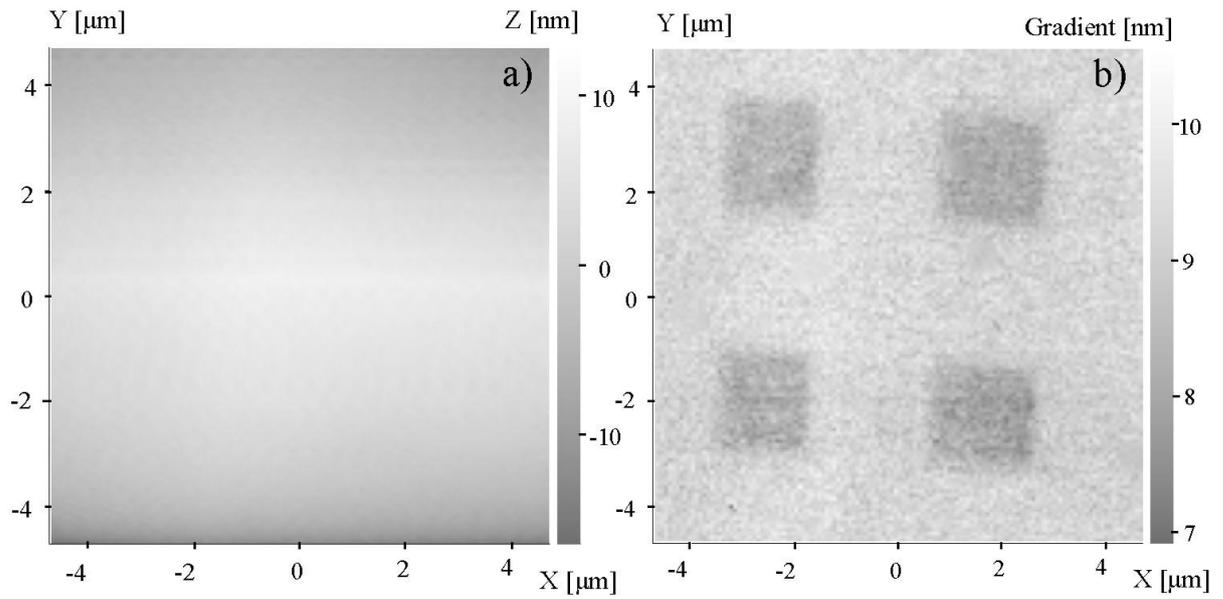

Figure 7. AFM image of the SiO$_2$/Si substrate affected by e-beam irradiation: the surface topography (a) and gradient (b).

But AFM scanning in gradient mode demonstrates contrast between exposed and unexposed areas (Figure 7b), which is likely related to the charge accumulated in the SiO$_2$-layer, that influences the cantilever. It should be noted that AFM investigation was made several days later after the electron beam exposure which suggests that the charge in the dielectric layer can persists for a long time. All this suggests that the charge in the silicon dioxide substrate can really affect the deposition rate of graphenic carbon thin films.

As mentioned in section 1, a detailed mechanism of graphene deposition on a non-metal surface is not yet clear. On the other hand, it is quite evident that the decomposition of carbon-containing molecules on the thermodynamically stable substrate occurs more readily than in the gas phase. As opposed to hydrocarbons, conventionally used as a precursor for graphene growth, alcohol molecules possess a polarity due to the OH-group presence. Perhaps these polar molecules can be involved in the field of persisted charge, and the concentration of adsorbed molecules on irradiated areas can increase. Accordingly, the rate of graphenic film formation increases too. Note that the above arguments are speculative and require experimental verification.

Thus, the graphenic carbon thin films growth is affected by the preliminary e-beam irradiation of the substrate and the rates of film deposition on the e-beam-exposed sites and on the unexposed areas are different. This effect can be assumed to arise as a result of charging the substrate, but this assumption requires experimental and theoretical studies of the mechanism.

## 4. Summary

The reported results show that nanographene films can be selectively grown on non-metal surface. For the silicon dioxide substrates we observed the influence of the pre-deposition electrons exposure upon the growth of graphenic carbon thin films. The conditions for the selective growth were found for the first time by varying the reaction time and water content in the initial mix. We are the first to find these conditions. Moreover, the selectively grown films are conductive and have linear I-U characteristic. The most probable mechanisms of the disclosed effect were investigated. Using Raman spectroscopy and atomic force microscopy we excluded the influence of the following factors on our experiments: formation of intermediate carbides, surface carbon contamination, formation of nanoclusters and defects which can act as growth centers. The result led us to suggest that the influence on the growth process consists in charging the substrate, so that the accumulated charge can affect the rate of graphenic films deposition.

The proposed method is highly promising for direct production of microstructures without post-synthesis lithography. Moreover, the CVD synthesis potentialities give possibilities for the selective graphenic carbon thin films growth on different dielectric substrates (quartz, sapphire, mica). So, more detailed experimental and theoretical study of the deposition process in relation to an e-beam pre-exposure and synthesis conditions as well as direct creature microstructure at the pre-synthesis step will be the subject of future investigations.

**Acknowledgements**

We acknowledge Dr. A.A. Svintsov (IMT RAS), Dr. S.I. Zaitsev (IMT RAS) and Dr. V.I. Korepanov (IMT RAS) for discussions.